\documentclass[runningheads]{llncs}
\usepackage{graphicx}
\usepackage{amsmath,amsfonts}
\usepackage{algorithmic}
\usepackage{graphicx}
\usepackage{textcomp}
\usepackage{xcolor}
\usepackage{multirow}
\usepackage{booktabs}
\usepackage{comment}
\usepackage{subfigure}
\usepackage{amsmath}
\usepackage{xspace}
\usepackage{balance}
\usepackage{todonotes}
\usepackage[hidelinks]{hyperref}
\usepackage{caption} 
\usepackage{subcaption}
\usepackage{pifont}
\usepackage{tcolorbox}

\begin{document}
\title{Transient Adversarial 3D Projection Attacks on Object Detection in Autonomous Driving}
\titlerunning{Transient Adversarial 3D Projection Attack}

\author{Ce Zhou\inst{1} \and
Qiben Yan\inst{1}\thanks{Corresponding author.} \and
Sijia Liu\inst{1}}
\authorrunning{Zhou et al.}
\institute{Michigan State University, East Lansing MI 48823, USA
\email{\{zhouce,qyan,liusiji5\}@msu.edu}}
\maketitle              % typeset the header of the contribution
\begin{abstract}
Object detection is a crucial task in autonomous driving. While existing research has proposed various attacks on object detection, such as those using adversarial patches or stickers, the exploration of projection attacks on 3D surfaces remains largely unexplored. Compared to adversarial patches or stickers, which have fixed adversarial patterns, projection attacks allow for transient modifications to these patterns, enabling a more flexible attack. In this paper, we introduce an adversarial 3D projection attack specifically targeting object detection in autonomous driving scenarios. We frame the attack formulation as an optimization problem, utilizing a combination of color mapping and geometric transformation models. Our results demonstrate the effectiveness of the proposed attack in deceiving YOLOv3 and Mask R-CNN in physical settings. Evaluations conducted in an indoor environment show an attack success rate of up to 100\% under low ambient light conditions, highlighting the potential damage of our attack in real-world driving scenarios.

\end{abstract}
\section{Introduction}

Object detection is a crucial task of autonomous driving systems, playing a pivotal role in ensuring the safety of human life. Accurate and reliable object detection enables autonomous vehicles (AVs) to perceive and respond to their environment, and to recognize and track objects such as pedestrians, vehicles, and obstacles. This capability is essential for making informed decisions and taking appropriate actions to navigate through complex and dynamic traffic scenarios. 

However, it is a well-recognized challenge that Deep Neural Networks (DNNs) based object detectors are vulnerable to adversarial perturbations. 
Previous research has designed various adversarial examples (AEs) within the digital domain to attack object detection models. To ensure physical realizability, existing studies create adversarial patches or stickers~\cite{eykholt2018robust,song2018physical,sharif2016accessorize} as AEs that take real-world environmental conditions into account, including diverse viewing distances or angles. 
%However, this type of attack is easy to detect by countermeasures due to their ~\cite{chou2020sentinet}. Besides,
However, patch-based attacks are indiscriminate toward all AVs, lacking flexibility in the physical world. This type of attack usually cannot bypass the defense mechanisms~\cite{chou2020sentinet}, as these adversarial attacks would largely perturb the model input. To overcome these limitations, the researchers proposed projection attacks to project the short-lived perturbations onto the target object to render it undetectable by the object detector~\cite{lovisotto2021slap,gnanasambandam2021optical}. These adversarial perturbations are crafted for use on 2D surfaces or relatively flat small 3D surfaces. However, most objects have curved or uneven large 3D surfaces, making the existing perturbations ineffective as their shapes can become distorted when projected onto a 3D surface.
%resulting in insufficiency to address real-world challenges by projecting these patches.
\begin{figure}[t]
\centering
	\includegraphics[width=1\columnwidth]{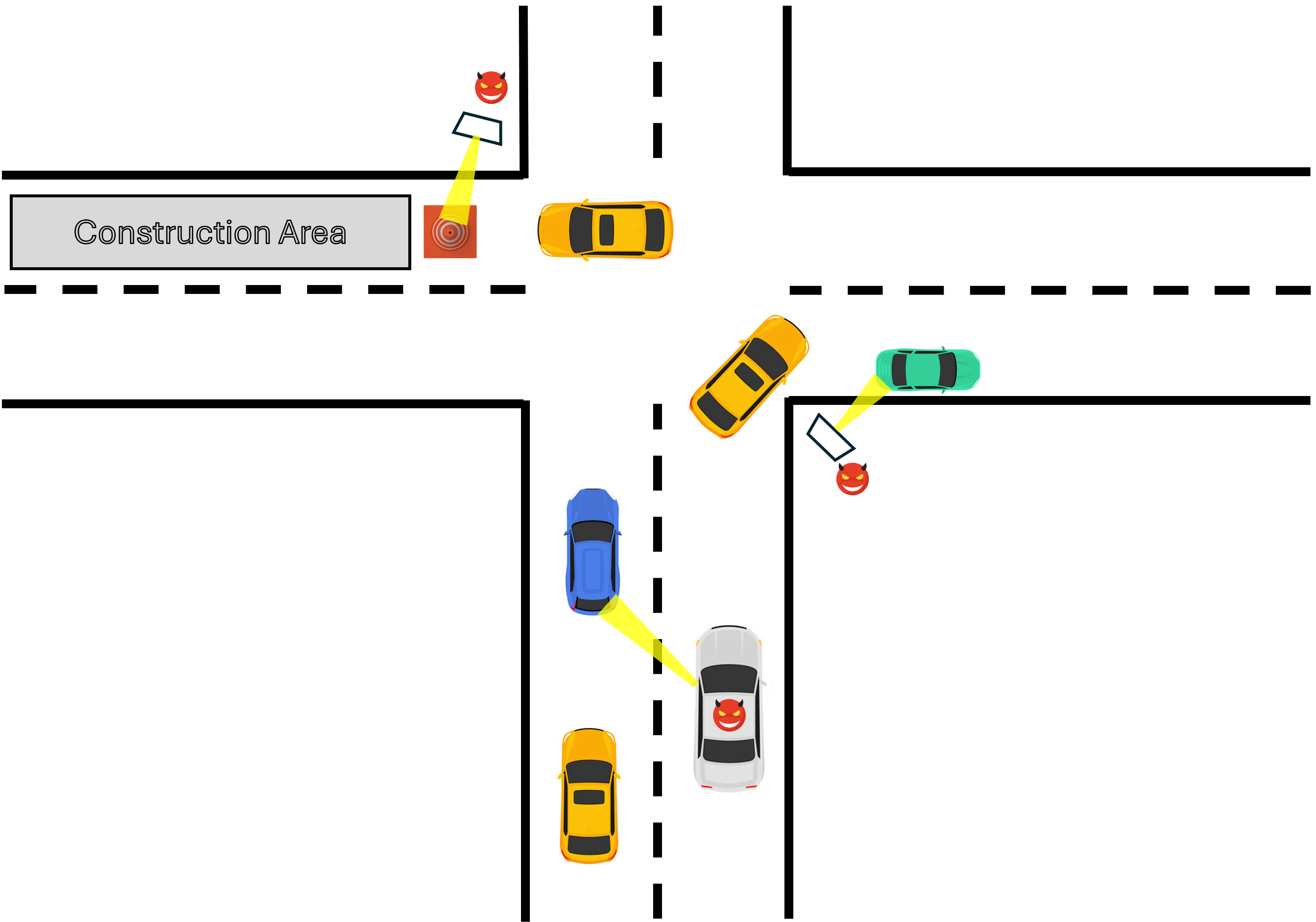}

 \caption{The attack scenarios of 3D projection attacks: \ding{182} The blue car and the white car are driving relatively static. \ding{183} The green car parks at the roadside. \ding{184} The traffic cone is placed at the entrance of the road construction area. The attackers position themselves either on the roadside or in another car. From these vantage points, the attacker can project an adversarial patch (projection light beams in yellow) onto the target vehicles or traffic cones, to render it undetectable by the victim AV (vehicles in yellow). This strategy could potentially lead to a collision between the victim AV and the target vehicles, or lead the victim vehicle into a dangerous road construction area.}
	\label{fig:attack_senarios}
\end{figure}

In this paper, we propose an adversarial 3D projection attack against object detection in autonomous driving scenarios. The proposed attack can work on a curved or uneven 3D surface with varying viewing angles. To the best of our knowledge, this is the first physically-realizable 3D projection attack towards AVs. 
%, which fills the current research gap. 
We illustrate three attack scenarios in Fig.~\ref{fig:attack_senarios}. The attack goal is to make the target vehicle undetectable by the victim AV, which could potentially lead to a collision between the victim AV and the target objects. In the first attack scenario, the attacker is driving the white car at the same driving speed as the blue car. The attacker projects the adversarial patch onto the surface of the blue car, causing it to disappear from the victim AV's object detector. In the second attack scenario, the attacker projects an adversarial patch onto a green car parked on the roadside. The victim AV fails to detect the green car, potentially leading to a crash. The third attack scenario is to project an adversarial patch onto the traffic cone to render it undetected by the victim AV's object detector. This could result in the victim AV driving into a dangerous road construction area. 

Several research challenges arise in implementing the proposed attack. When projecting the image onto the object's surface, the resulting images are influenced by various factors, including surface color, material composition, ambient light, projection distortion, viewing angle, etc. First, we need to solve the research question: ``\textit{How to map the projected image to the resulting image on the object surface to preserve the intended color schemes?}''. Second, the 3D surface is always curved or uneven, which leads to the distortion of the projected adversarial patch. Thus, the research question is: ``\textit{How to transform the projected patch onto the 3D surface to ensure its effectiveness?}''. Third, when projecting the attack pattern onto the target vehicle, the attacker and the target vehicle may both move. So, another research question is: ``\textit{How to ensure the attack's robustness in a dynamic environment?}''. 

To address these challenges, we employ a color mapping model to map the projection color onto the 3D attack surface. Additionally, we introduce a geometric transformation model that employs the Thin Plate Spine (TPS) algorithm to perform the geometric transformation from the projected patch to the 3D attack surface. To enhance attack robustness, we enhance the training dataset by applying different transformations, i.e., Expectation Over Transformation (EOT), to input images captured with various environmental factors, such as view angles, distances, etc.

We conduct experimental evaluations in an indoor environment with a model vehicle and present a video demo\footnote{\url{https://youtu.be/8RbDpAAmsjs}} to demonstrate the 3D projection attack. We list our contributions as follows:
\begin{itemize}
    \item We are the first to propose a transient and physically-realizable projection attack by projecting an adversarial patch on a 3D object surface in autonomous driving scenarios.
    \item We formulate the adversarial patch generation as an optimization problem by combining a color projection model and a geometric transformation model with the target object detection network. We enhance the training data by considering various environmental conditions and different viewing angles of the adversarial patch, which ensures the effectiveness and robustness of the attack.
    \item We evaluate our attack on two object detection models using a 1/10 scale remote control (RC) car in an indoor environment. We run the experiments under varying attack distances, angles, and ambient light conditions. Evaluations show an attack success rate of up to 100\% under low ambient light conditions, demonstrating the potentially damaging consequences of the proposed attack in real-world driving scenarios.
\end{itemize}

\section{Background}

We first introduce the background knowledge of projector technology and object detection. Then, we present existing literature related to physically-realizable AEs.

\subsection{LCD Projector Technology} 

Liquid Crystal Display (LCD) projectors, widely employed in business presentations, classrooms, and home entertainment, utilize liquid crystal technology to project images. The fundamental concept involves splitting white light emitted by a lamp into its red, green, and blue components through dichroic mirrors. These mirrors reflect specific wavelengths while allowing others to pass through. Consequently, the white light transforms into the primary colors of red, green, and blue, forming the basis for deriving all other colors~\cite{LCD_Scott_Wilkinson}. 

Lumens of the projector measure the brightness of a projector, indicating the total amount of light it emits. A higher lumen rating in LCD projectors enhances image visibility, especially in well-lit surroundings. Indoor projectors typically fall within the 2,000 to 3,000 lumens range for emitted light, whereas some outdoor projectors can achieve super-high lumens. 
%The existing ambient light on the projection surface significantly influences the formed image, with brighter ambient light resulting in less visibility due to diminished contrast and a more limited color range~\cite{lovisotto2021slap}.

The throwing ratio of an LCD projector is defined as the distance between the projector and the screen relative to the width of the projected image. Lux is the unit for one lumen per square meter. Therefore, for a determined projecting distance, the throwing ratio influences the size and brightness of the projected image.
%A lower throwing ratio results in a shorter throw distance, ideal for smaller rooms or spaces with limited projection space. Conversely, a higher throwing ratio is suitable for larger venues, allowing flexibility in projector placement. Striking the right balance between lumens and throwing ratio is essential for achieving the desired image size and quality.

%Resolution and Color Accuracy:
%While lumens and throwing ratio are pivotal, resolution and color accuracy are equally important aspects of LCD projector technology. Higher resolutions, such as Full HD (1920 x 1080) or 4K (3840 x 2160), contribute to sharper and more detailed images, enhancing the overall viewing experience. Additionally, advancements in color processing technologies, such as 3LCD, ensure accurate color reproduction, which is crucial for presentations where color fidelity is paramount.

When we project a 2D image onto a 3D surface using a projector, the appearance of the image can be affected by the shape and texture of the surface. On smooth surfaces, the image is likely to be more uniform. However, any irregularities in the surface might cause slight distortions or uneven brightness. Textured surfaces can cause the projected image to look uneven, as the texture may interfere with the clarity of the image. If the surface is flat, the projected image will appear as it does on a flat screen, whereas, on curved 3D surfaces, the image may appear distorted, especially towards the edges. The amount of distortion depends on the degree and nature of the curvature. In a real-world attack scenario, both vehicles and road objects, such as traffic cones, have 3D curved surfaces, making existing adversarial attacks ineffective. This paper aims to develop a new projection attack that ensures attack effectiveness on these 3D surfaces. 

\subsection{Object Detection}

Object detection, a computer technology within the domain of computer vision and image processing, is centered on identifying instances of semantic objects belonging to specific classes, such as pedestrians and cars, in images and videos~\cite{Object_detection_wiki}. 
%In this paper, we will focus on Yolov3~\cite{redmon2018yolov3}, which is adoped in our evaluation. 
In this paper, we focus on %Yolov5~\cite{Jocher_YOLOv5_by_Ultralytics_2020}, 
Mask-RCNN~\cite{ren2015faster} and Yolov3~\cite{redmon2018yolov3}, as they are representative object detectors widely used in computer vision research. 

Faster-RCNN is an evolution of the initial R-CNN object detector network~\cite{girshick2014rich}. Utilizing a two-stage detection method, the first stage is to generate region proposals, and the second stage is to predict labels for the proposals.
%Faster-RCNN employs an initial network to generate region proposals, and a second network predicts labels for these proposals. 
Additionally, Mask-RCNN~\cite{he2017mask} expands Faster-RCNN to incorporate instance segmentation, providing precise pixel-level masks alongside accurate object detection, making it ideal for applications requiring detailed object boundaries. Its flexible architecture and strong performance on benchmark datasets make it a versatile choice. 

Yolo object detectors are designed to detect objects in real-time, making them highly suitable for various applications~\cite{yolo_Ritesh_Kanjee}. Yolov3 is a one-stage detector utilizing a single convolutional neural network (CNN), which incorporates a backbone network to compute feature maps for each square-grid cell in the input image. It employs three different grid sizes to enhance the accuracy of detecting smaller objects. Its unified architecture simplifies implementation, and its ability to perform multi-scale detection ensures robustness across various object sizes. Both Yolov3 and Mask-RCNN employ non-maximum suppression in post-processing to eliminate highly overlapped redundant boxes.

%YOLOv8, developed by Ultralytics, is considered the latest and most advanced version of YOLO models. Known for its fast processing speed, high accuracy, and ability to detect multiple objects in a single image, YOLOv8 features a new backbone network, facilitating easy performance comparison with older YOLO models. It also incorporates a new loss function and an anchor-free detection head.

\subsection{Physical Adversarial Examples}

Athalye et al.~\cite{athalye2018synthesizing} introduce the EOT framework, focusing on crafting adversarial perturbations resilient to random linear transformations. Although this approach uses a 3D printer to print out the 3D object with the AE, which can successfully deceive the image classifiers, it is not flexible in the context of dynamic driving scenarios. 

%The success of this approach is demonstrated by perturbing the texture of a digital 3D object, later printed using a color 3D printer. However, this attack targets a single object, not addressing the complexity of real-world scenes.

Other studies~\cite{eykholt2018robust,song2018physical,zhao2019seeing} design robust physical perturbations, like stickers or patches, for traffic signs. These perturbations are proven resilient to changes when reproduced in the physical world, such as alterations in distance and viewing angle. Sharif et al.~\cite{sharif2016accessorize} showcase the feasibility of physical AEs for face recognition using colored eye-glass frames and enhance the perturbation realizability in the presence of input noise. Xu et al.~\cite{xu2020adversarial,fan2021generating} design a patch on a non-rigid object (i.e., T-shirt) to make the person undetected in the object detector. However, such patch-based attacks also lack the flexibility in a complex autonomous driving scenario.

Patches have drawbacks that can be mitigated by the projectors, particularly due to the short-lived and dynamic nature of projections. This allows attackers to control and adjust the projection easily, targeting specific vehicles without leaving traces of the malicious attack. Nassi et al.~\cite{nassi2020phantom} project 2D images directly into the real world to deceive object detectors. Man et al.~\cite{man2020ghostimage} project optimized adversarial perturbations directly to the camera using the flare lens effect to deceive image classification. Similarly, Zhou et al.~\cite{zhou2022doublestar} exploit vulnerabilities in stereo depth estimation algorithms and use two projectors to deceive depth estimation results for AVs. In~\cite{lovisotto2021slap,gnanasambandam2021optical}, the researchers project the perturbation on the traffic sign to make it undetectable by the object detector. In these cases, adversarial perturbations are crafted for use on 2D surfaces, despite most objects having 3D surfaces, rendering flat patches ineffective on curved 3D object surfaces.

\section{Threat Model}

Our proposed attack targets an autonomous driving scenario, in which the decision-making relies exclusively on camera sensors for the AV, such as Tesla vehicles with Full Self-Driving features and Active Safety Features~\cite{Autopilot_and_full_self}. The AV utilizes these cameras to identify and monitor 3D objects within the driving scene, including other moving vehicles.

\textbf{Attack Goal.} The adversaries aim to cause a traffic accident by projecting a transient adversarial patch onto a 3D object, making it undetectable by a DNN-based object detector model that processes the camera feeds within the AV. For instance, the lack of detection may cause the surrounding vehicles to collide with the victim AV without engaging the brakes.

\textbf{Attacker's Capabilities.} Our attack is a \textit{white-box, hiding attack}. We consider a scenario where an attacker seeks to promptly generate and project an attack patch onto an existing real-world 3D object to \emph{hide} it from an object detection system employing a DNN model. During this timeframe, every captured frame of the scene is anticipated to yield a deceptive object detection outcome. We assume the attacker possesses comprehensive knowledge of the target DNN, including the model parameters and architecture, while remaining unaware of the technical specifications of the cameras. Additionally, the attacker is assumed to have physical proximity to the target 3D objects, such as other vehicles.
This proximity enables the attacker to capture images of the target 3D objects for adversarial patch training and facilitates the projection of the attack patch into the real world.
 We also assume that the attacker has the capability to purchase or locate an identical object as a target for the projection attack, thereby gathering all the required data for the attack preparation (i.e., adversarial patch generation).

\section{Attack Design}\label{sec:attack_design}

We first present an overview of the attack pipeline, and then we illustrate the design details regarding the color mapping model, geometric transformation model, adversarial patch generation, and training data enhancement. 

\begin{figure}[t]
\centering
	\includegraphics[width=1\textwidth]{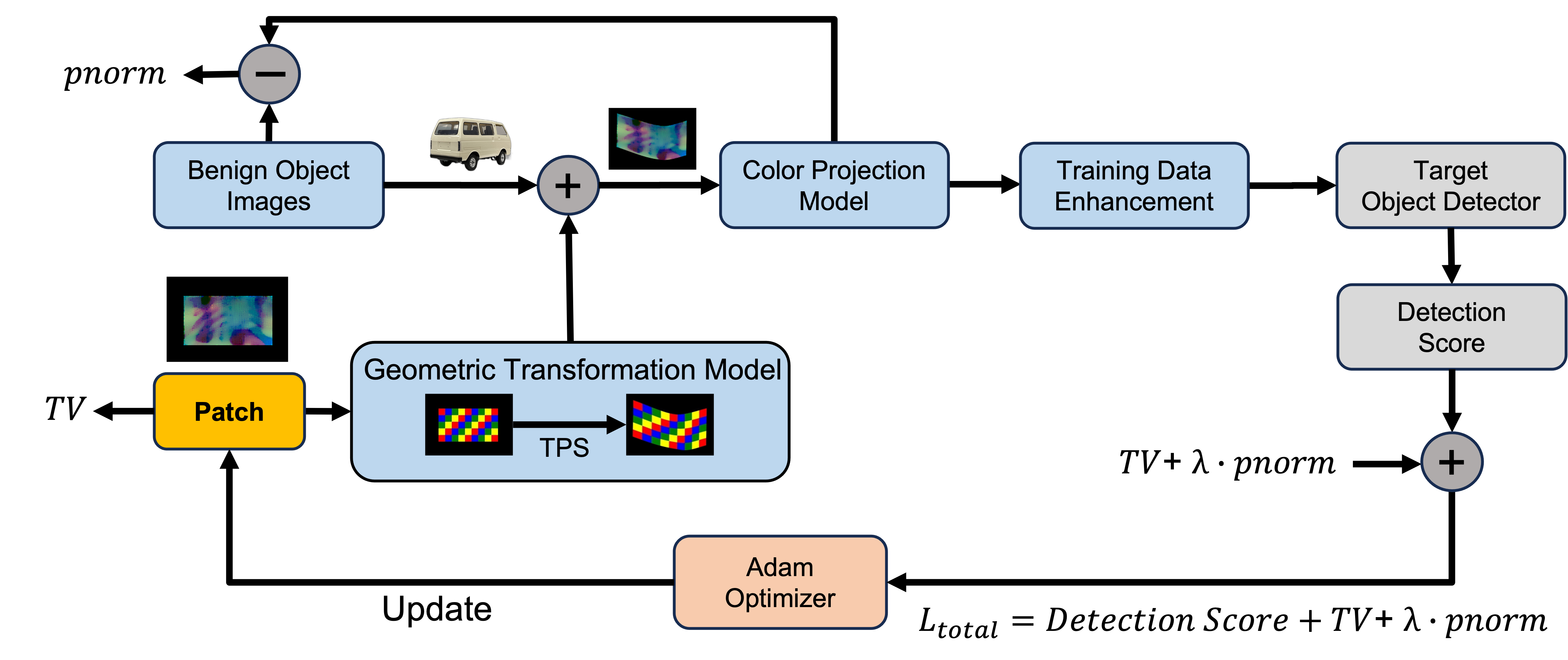}
	\caption{Overview of the adversarial patch generation pipeline.}
	\label{fig:attack_pipline}
\end{figure}

\subsection{Attack Overview}

%Figure~\ref{fig:attack_pipline} shows the attack pipeline. We optimize the adversarial patch that undergoes the color mapping model to minimize the target object detection score on a specific DNN-based object detector. The optimization process is carried out across the augmented training dataset, which consists of a collection of attacked images incorporating EOT. 

Fig.~\ref{fig:attack_pipline} shows the adversarial patch generation pipeline. The goal is to optimize the adversarial patch using a geometric transformation model and a color mapping model to minimize the target object detection score on a specific DNN-based object detector. The adversarial patch undergoes geometric transformation, and then the color mapping model is applied to simulate the patch's appearance on the target object.
 The geometric transformation model simulates the distortion on the 3D object surface, applying the transformation on the patch based on the 3D structure of the target object. The color mapping model simulates the projected color on the target surface.
The transformed patch and the benign target object are combined through the color projection model to generate the patched final object. We then perform training data enhancement to place the patched target object in different backgrounds to enhance robustness in real-world driving scenarios. Finally, the optimization process is conducted iteratively across the enhanced training dataset to generate the adversarial patch (for projection) by minimizing the loss function.

\subsection{Color Mapping Model}

To create physically-realizable perturbations, prior studies~\cite{sharif2016accessorize,eykholt2018robust,song2018physical} typically leverage the non-printability score (NPS) to characterize printable colors by printers. In our attack scenarios, projecting different colors onto the target 3D object surface yields results influenced by multiple factors, such as projection strength, ambient light, projection surface, etc. Some previous investigations on the projection-based physical attacks \cite{lovisotto2021slap,gnanasambandam2021optical} propose color mapping models to simulate the color overlays. 
%The achievable color spectrum is notably smaller compared to the spectrum available to printed patches due to these factors (e.g., a patch can be black or white, while most projections on a stop sign will result in red-ish images). 
The attainable range of colors is significantly reduced in comparison to the spectrum accessible to printed patches, primarily because of the color and materials of the target surface. 
%Therefore, we formulate the color mapping model as follows~\cite{lovisotto2021slap}.

Our objective in developing the color mapping model is to establish a mapping that associates colors from the projected image onto the target 3D surface. Therefore, we formulate our color mapping model $\mathcal{P}_c$ as follows:
\begin{equation}
   \mathcal{P}_c(\theta_{p},S,P)=O,
   \label{equ1}
\end{equation}
where $\theta_{p}$ is the model parameters, $S$ is the 3D target surface, $P$ is the projected image, and $O$ is the color overlays captured by the camera on the target 3D surface. 

%Subsequently, we gather data to fit the color mapping model, aiming to approximate the resultant output color for given projected images and target 3D surfaces. The fundamental approach involves projecting various colors with RGB values ranging from 0 to 255 onto the target 3D surface, capturing images each time to document the outcomes. However, collecting data for all possible colors ($255^{3}$ different colors) in real-world scenarios is highly time-consuming, and many of these colors may appear very similar. Therefore, we opt for a more efficient approach by sampling with a step size of 127, resulting in $3^{3}=27$ distinct color mappings denoted as $(S, P_i) \Rightarrow O_i$, where $i$ represents each color index. 
Subsequently, we collect projected data to train the color mapping model, with the goal of approximating the resulting output color for specific projected images and target 3D surfaces. The fundamental approach involves projecting various colors with RGB values ranging from 0 to 255 onto the target object surface, capturing images each time to document the outcomes. The distinct color mappings are denoted as $(S, P_i) \Rightarrow O_i$, where $i$ represents each color index. 

%To make sure the data collection is accurate enough, we need to consider two factors~\cite{lovisotto2021slap}: sensitivity of the light sensor (ISO~\cite{mancuso2001introduction}) under different light conditions and smoothing over-time effect in the sensor readings while recording the video. To mitigate the influence of these two factors, we capture 10 consecutive frames for each color. We intersperse each projected color with 10 frames of no mapping and calculate the median pixel value for each corresponding pixel across these frames. This process is conducted with a static camera setup. The resulting median values are utilized as our final images. 

To train the color mapping model, we utilize triples $(S, P_i, O_i)$ to fit a neural network consisting of two hidden layers with ReLU activation. Then, Eq.~(\ref{equ1}) is transformed into an optimization problem with the following loss function:
\begin{equation}
    Loss_{\mathcal{P}_c}=\underset{\theta_p}{\operatorname{arg min}} \sum_{\forall i}||\mathcal{P}_c({S,P_{i}})-O_{i}||_{1},
\end{equation}
where $||\cdot||_1$ denotes the L1 norm. We optimize the network parameters $\theta_p$ using gradient descent and the Adam optimizer to derive the color mapping model.

\subsection{Geometric Transformation Model}

\begin{figure}[t]
    \centering  
    \subfigure[Color board patch.]{\includegraphics[width=0.26\textwidth]{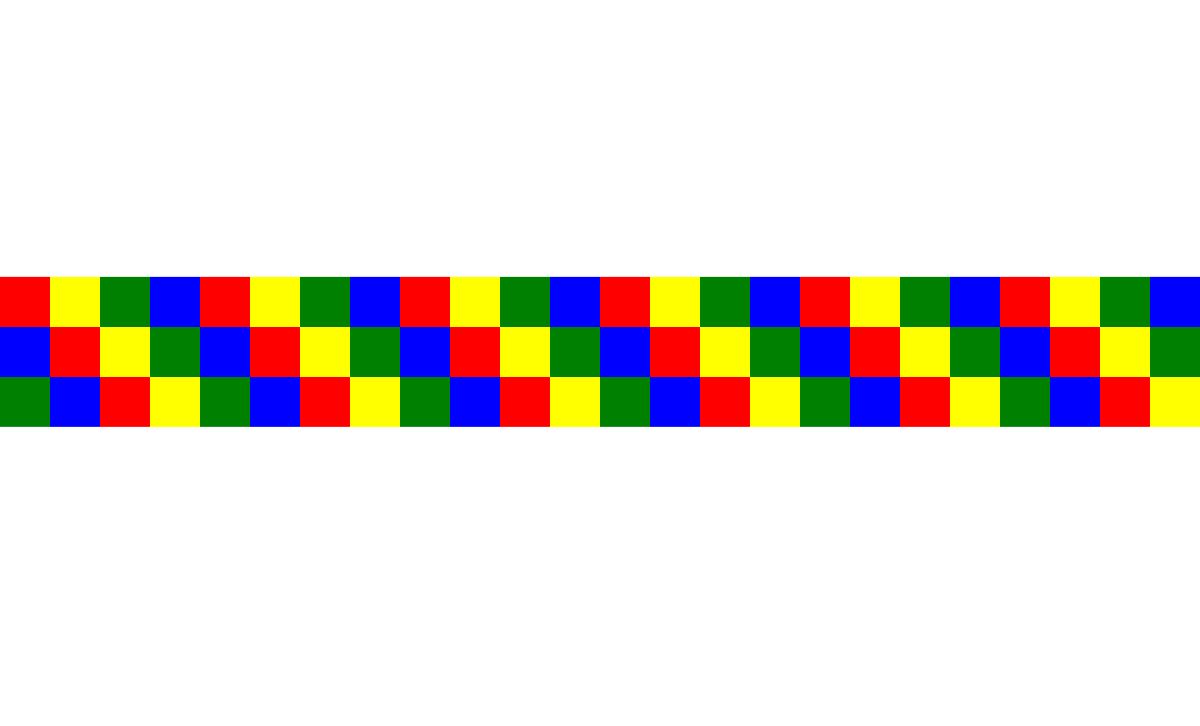}} 
    \subfigure[Viewing angle 1.]{\includegraphics[width=0.33\textwidth]{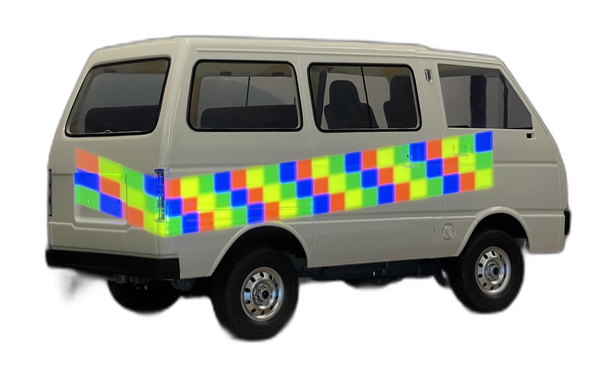}} 
    \subfigure[Viewing angle 2.]{\includegraphics[width=0.33\textwidth]{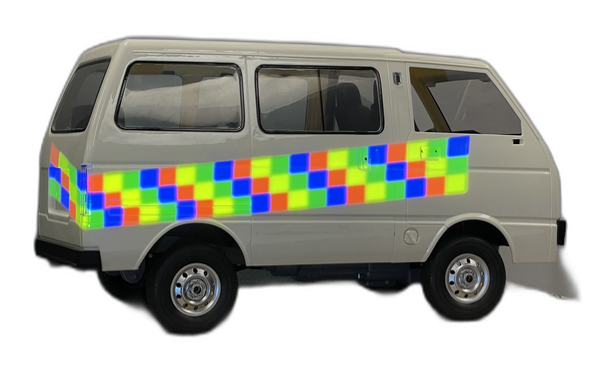}}     
    \caption{Examples of two viewing angles (b)(c) of the same (a) color board patch projection on the same vehicle.}
    \label{fig:different_angle}
\end{figure}

The EOT process~\cite{athalye2018synthesizing} involves using images with various transformations, encompassing scaling, translation, rotation, brightness, additive Gaussian noise, etc. This collection aids in creating a single perturbation effective in deceiving the target neural network across diverse views. However, this approach may not adequately capture the projected image warping on a 3D object surface, particularly when viewed from different angles. For instance, in Fig.~\ref{fig:different_angle}, the projected patch, i.e., the color board, on the 3D surface is unevenly distorted on the car's surface. Thus, it is imperative to consider this viewing angle aspect when modeling the transformation.

Different viewing angles or relative movements between the projected 3D surface and the projector can result in varying distortions in each video frame. To address this, we incorporate TPS~\cite{bookstein1989principal} to model the projection distortion on the 3D surface. TPS has been widely employed as a non-rigid transformation model in prior studies~\cite{jaderberg2015spatial,xu2020adversarial,fan2021generating}. TPS transformation is a mathematical model used for non-linear spatial transformations, often applied in image warping and deformation. It is particularly useful in morphing one image into another while preserving local structures. TPS transformation is defined as a combination of affine, rigid, and non-rigid components. Our exploration will reveal that the non-rigid warping aspect of TPS proves to be an effective means of modeling projection distortion on 3D surfaces in learning adversarial patterns.

TPS involves learning a parametric deformation mapping that describes the displacement of each pixel from an original image \(x\) to a target image \(z\) using a set of control points with predefined positions. Suppose that the control points in the source image are \((x_i, y_i)\) and their corresponding positions in the target image are \((z_i, w_i)\). The TPS transformation \(\mathcal{P}_g\) is defined as:
\begin{equation}
    \mathcal{P}_g(x, y) = f(x, y) + \sum_{i=1}^{N} w_i \cdot \phi(r_i),
    \label{equ:tps}
\end{equation}
where \( f(x, y) \) is an affine function, and \( \phi(r_i) \) is a radial basis function defined as \( \phi(r) = r^2 \cdot \log(r) \). The Euclidean distance \(r_i\) between the point \((x, y)\) and the control point \((x_i, y_i)\) is given by:
\[
r_i = \sqrt{(x - x_i)^2 + (y - y_i)^2}.
\]

With the determined transformation parameters, it also allows the reverse transformation from the target image \(z\) back to the original image \(x\). The reverse transformation is given by:
\[
\mathcal{P}_g^{-1}(z) = \sum_{i=1}^{N} w_i \cdot \phi(r_i).
\]

Please note that Eq.~(\ref{equ:tps}) does not explicitly define \( z_i \). Instead, it uses the notation \( w_i \cdot \phi(r_i) \) to represent the non-rigid component of the transformation. \( z_i \) is implicitly part of the control points \((z_i, w_i)\) in the target image but does not appear directly in the equation. In other words, \( \sum_{i=1}^{N} w_i \cdot \phi(r_i) \) already incorporates the influence of these control points in the transformation.

Implementing TPS to design a 3D adversarial projection patch is challenging due to the difficulty in determining the control points on both the projected and captured target surface images. To tackle this challenge, we project a checkerboard onto the target object and identify the intersection points between adjacent grid regions as the control points by manually marking them. We then scale the projected image and mark the control points using the same procedure.

\subsection{Adversarial Patch Generation}

Our approach for generating the adversarial projection pattern involves combining the color projection model and the geometric transformation model with the target network. We employ gradient descent along both dimensions to optimize the projected image. Our objective function is defined as:
\begin{align}
    \underset{\delta}{\operatorname{arg min}} J(f_\theta(b+\mathcal{P}(x,\delta))), &&\textrm{s.t. } 0\leq\delta\leq1,
    \label{equ:adv1}
\end{align}
where $\delta$ is the projected image, $x$ is the target object image, $J$ is the detection loss, $f_\theta$ is the target object detector, and $b$ is the input image background. $\mathcal{P}$ is the projection model which is defined as:
\begin{equation}
    \mathcal{P}(x,\delta)=(1-M)\cdot x+\mathcal{P}_c(M\cdot x,\delta_g),
    \label{equ2}
\end{equation}
where $M$ is the mask of the patch on the target object image, and $\delta_g$ is the patch on the target object image. 
Here, $M=\mathcal{P}_g(S_\delta)$, and $\delta_g=\mathcal{P}_g(\delta)$, respectively. $S_\delta$ denotes the shape of the projected image. We then rewrite Eq.~(\ref{equ2}) as:
\begin{equation}
    \mathcal{P}(x,\delta)=(1-\mathcal{P}_g(S_\delta))\cdot x+\mathcal{P}_c(\mathcal{P}_g(S_\delta)\cdot x,\mathcal{P}_g(S_\delta)).  
    \label{equ3}
\end{equation}
We predefine the projected patch shape. The projected patch goes through the geometric transformation model and color mapping model before it is applied to the input image. We mask the attack area on the object to ensure the perturbations only apply to the region of interest (ROI) on the 3D object surface.

To improve the practical feasibility of the projection, we impose a constant grid granularity of $n\times n$ cells on the projection, similar to the procedure in~\cite{lovisotto2021slap}. This guarantees that each cell consists of pixels with identical colors, promoting uniform projections across various viewing distances of the target object. Additionally, we integrate the total variation in the loss function. This inclusion is intended to alleviate the effects of camera smoothing and/or blurring on the overall projection~\cite{mahendran2015understanding}.

To simplify the flowing of gradients when backpropagating, we substitute $\delta$ with a new variable $w$ such that:
\begin{equation}
    u=\frac{\tanh\delta}{2}+0.5.
\end{equation}
Since $\delta$ is bounded in [0,1], therefore, $u$ is bounded in [-1,1], which leads to faster convergence in the optimization~\cite{lovisotto2021slap}.

Now, we can update our loss function Eq.~(\ref{equ:adv1}) to constrain the amount of perturbations as:
\begin{align}
    \underset{\delta}{\operatorname{arg min}} J(f_\theta(b+\mathcal{P}(x,u))) +\lambda||\mathcal{P}(x,u)-x||_p +TV(u) \nonumber\\    \textrm{s.t. } -1\leq u\leq1, 
\end{align}
where $\lambda$ is a parameter utilized to regulate the significance of the p-norm $||\cdot||_p$ and $TV$ represents the total variation. The loss $J$ is based on bounding boxes \(b \in B\) outputted from the target object detection network. Each bounding box has a confidence score indicating the probability of containing an object of class \(j\), denoted as $p_j^{(b)}$. The loss $J$ is then defined as the sum of the confidence scores for the target object, expressed as \(\sum_{b \in B} p_j^{(b)}\).

\subsection{Training Data Enhancement}

Generating AEs that effectively function in autonomous driving scenarios necessitates the consideration of various environmental conditions, such as different viewing angles, distances, rotation, and brightness. To craft a robust adversarial patch, the optimization process needs to incorporate different input transformations. We adopt EOT to generate a set of training images synthetically. Our final loss is formulated as follows:
\begin{align}
    \underset{\delta}{\operatorname{arg min}_{x_q\sim X}} \mathbb{E}_{b_i\sim B,m_j\sim M}J(f_\theta(b+m_j\cdot \mathcal{P}(x_q,u))) \nonumber+\lambda||\mathcal{P}(x_q,u)-x_q||_p +TV(u), \nonumber\\\textrm{s.t. } -1\leq u\leq1. 
\end{align}
where $\mathcal{P}(x_q,\delta)=(1-\mathcal{P}_g^q(S_\delta))\cdot x_q+\nonumber\mathcal{P}_c(\mathcal{P}_g^q(S_\delta)\cdot x_q,\mathcal{P}_g^q(S_\delta))$. \(X\) represents a collection of input images with various viewing angles, \(B\) is a set of background images, and \(M\) denotes a number of linear transformations to the target object with the patch. \(\mathcal{P}_g^q\) denotes the geometric transformation model corresponding to the input object image \(x_q\).

\subsection{Overall Attack Process}
We summarize the detailed attack process of our attack step by step as follows:

$\bullet$ Step 1: The attackers determine the patch shape and obtain the parameters for the geometric transformation model. First, the attackers design a patch shape, such as a rectangle, to be used as input to the projector. They use a color board with the same shape as the patch to project it onto the surface of the target 3D object. They collect images of the object both with and without the projection. 
Finally, they select the control points for the TPS and derive the parameters for the geometric transformation model based on these control points.

$\bullet$ Step 2: The attackers collect data for the color projection model and train the model. They project different colors onto the target object and capture images of the object with these projections. This data is then fed into the color projection model for training.

$\bullet$ Step 3: The attackers generate the patch using the enhanced training dataset. The transformed adversarial patch and collected benign images are processed through the color mapping model and applied to the target objects. They then enhance the training dataset by incorporating EOT, placing the target objects with the adversarial patches in different backgrounds.

$\bullet$ Step 4: The attackers run the optimization process and update the patch. They conduct the optimization process across the enhanced training dataset using the Adam optimizer. The goal is to minimize the loss function, which includes the summation of the detection score, total variation, and p-norm of the patch. Finally, they update the patch accordingly based on the optimization results.

\section{Evaluation}\label{sec:attack_eva}

To evaluate the proposed attack, we conduct experiments on 3D projection attacks in the physical world. We investigate our attack under different environment settings to verify its feasibility and effectiveness.

\subsection{Experimental Setup}

\subsubsection{Attack Devices.}

\begin{figure}
\centering
	\includegraphics[width=0.65\columnwidth]{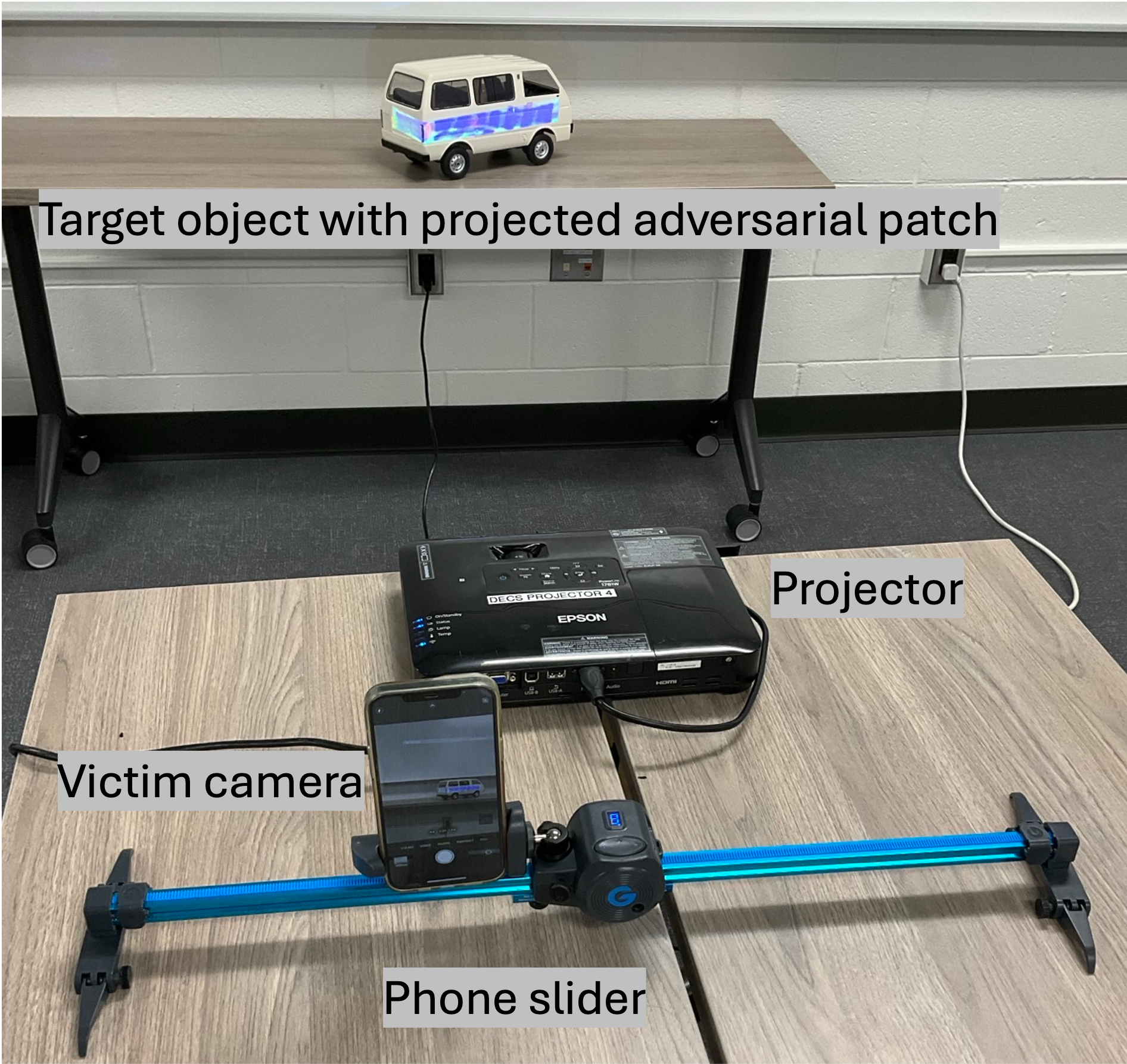}
	\caption{Attack setup. We use the projector to project the simulated adversarial patch on the target vehicle. The phone slider is used to collect videos when the victim's camera is moving.}
	\label{fig:attack_setup}
\end{figure}

Fig.~\ref{fig:attack_setup} illustrates the experimental setup. The target object is a 1/10 scale RC car~\cite{1_10_car}. For the projection, we use a PowerLite 1771W WXGA 3LCD Projector~\cite{PowerLite}, an indoor projector priced around \$740, which offers a maximum brightness of 3,000 lumens and a maximum resolution of $1,280 \times 800$ pixels. The projector has a throw ratio range of 1.04–1.26. The initial experiments were conducted in a lecture room, with the projector positioned 1.5 to 2.5 meters away from the target object. Given the 1/10 scale of the RC car, this setup simulates an attack distance of 15 to 25 meters in a real-world driving scenario.

To capture images and videos of the target object, we use the default camera on an iPhone 12 Pro Max. The iPhone is mounted on a phone slider~\cite{PhoneSlider} to capture videos from viewing angles ranging from $-20^\circ$ to $20^\circ$. These videos are recorded at various distances and angles while the projection is active, under ambient light conditions of 100 lux, 200 lux, and 500 lux.

Additionally, the data for the color mapping model are collected using the built-in webcam on an Alienware m15 R3 laptop~\cite{alienware}. The data for the geometric transformation model is collected by iPhone. The evaluation of the attack is conducted on the images and videos captured with the iPhone.

\subsubsection{Target Attack Models.} 
In our experiments, we evaluate our attack on YOLOv3 and Mask R-CNN object detectors. For YOLOv3, we utilize the Darknet-53 backbone~\cite{redmon2018yolov3}. For Mask R-CNN, we use ResNet-101 as a backbone along with a feature pyramid network for the region proposals~\cite{lin2017feature}. Both object detectors produce a set of bounding boxes with associated confidence scores for each output class. We establish the detection threshold at 0.6, meaning that an object is detected as a ``car'' if the confidence score is higher than 0.6.

\subsubsection{Evaluation Metrics.}
We feed each frame from the videos into the target attack models and count the instances where a ``car'' is detected at the output. We use \emph{object misdetection rate (OMDR)} as the main evaluation metric in the experiments. It is expressed as:
\[
\text{OMDR} = \frac{\text{The number of frames that do not output a class of car}}{\text{Total number of frames}}
\]

\subsubsection{Experimental Procedure.}

We follow these essential steps to conduct our experiments:
\begin{itemize}
    \item Designing the Attack Pattern: we determine an attack pattern shape (a rectangular shape) as the input to the projector. We design a color board based on this pattern shape and project it onto the target object. Using the iPhone, we capture images of the target object both with and without the projection to obtain the benign object images and the images needed to identify the target control points, respectively. With these collected data, we run the geometric transformation model to obtain the model parameters.
    
    \item Running the Color Mapping Procedure: we run the color mapping procedure to collect different color projection images using the webcam. The collected data is then used to train and construct a color mapping model. %This step follows a similar procedure as mentioned in \cite{lovisotto2021slap}.
    
    \item Generating the Adversarial Pattern: we use the color mapping model and the geometric transformation model to generate the adversarial pattern and obtain the optimized attack pattern to input into the projector.
    
    \item Launching Attacks and Recording: we project the trained patch onto the target vehicle and record a set of videos at different distances and angles under various ambient light conditions.
\end{itemize}

\begin{figure}[t]
    \centering    
    \subfigure[]{\includegraphics[width=0.23\textwidth]{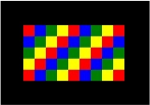}} 
    \subfigure[]{\includegraphics[width=0.23\textwidth]{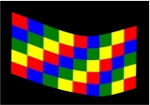}} 
    \subfigure[]{\includegraphics[width=0.23\textwidth]{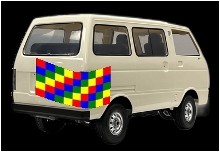}} 
    \subfigure[]{\includegraphics[width=0.23\textwidth]{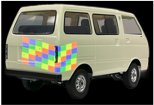}} 
    \subfigure[]{\includegraphics[width=0.23\textwidth]{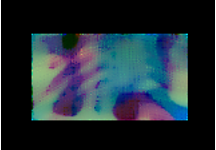}} 
    \subfigure[]{\includegraphics[width=0.23\textwidth]{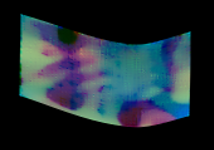}} 
    \subfigure[]{\includegraphics[width=0.23\textwidth]{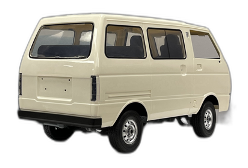}} 
    \subfigure[]{\includegraphics[width=0.23\textwidth]{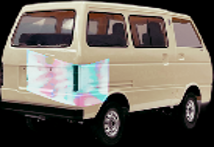}} 
    \caption{A visualization example of geometric transformation using TPS. (a) The color board used to collect TPS source control points; (b) The simulated color board after TPS transformation; (c) The simulated color board projected on the vehicle; (d) The projection of the color board on the real vehicle to collect TPS target control points; (e) The trained adversarial patch; (e) The simulated adversarial patch after geometric transformation; (g) The benign vehicle image; (h) The simulated patch projected on the vehicle.}
    \label{fig:TPS_visualization}
\end{figure}

\begin{figure}[htbp]
    \centering    
    \subfigure[Attack on Yolov3]{\includegraphics[width=0.45\textwidth]{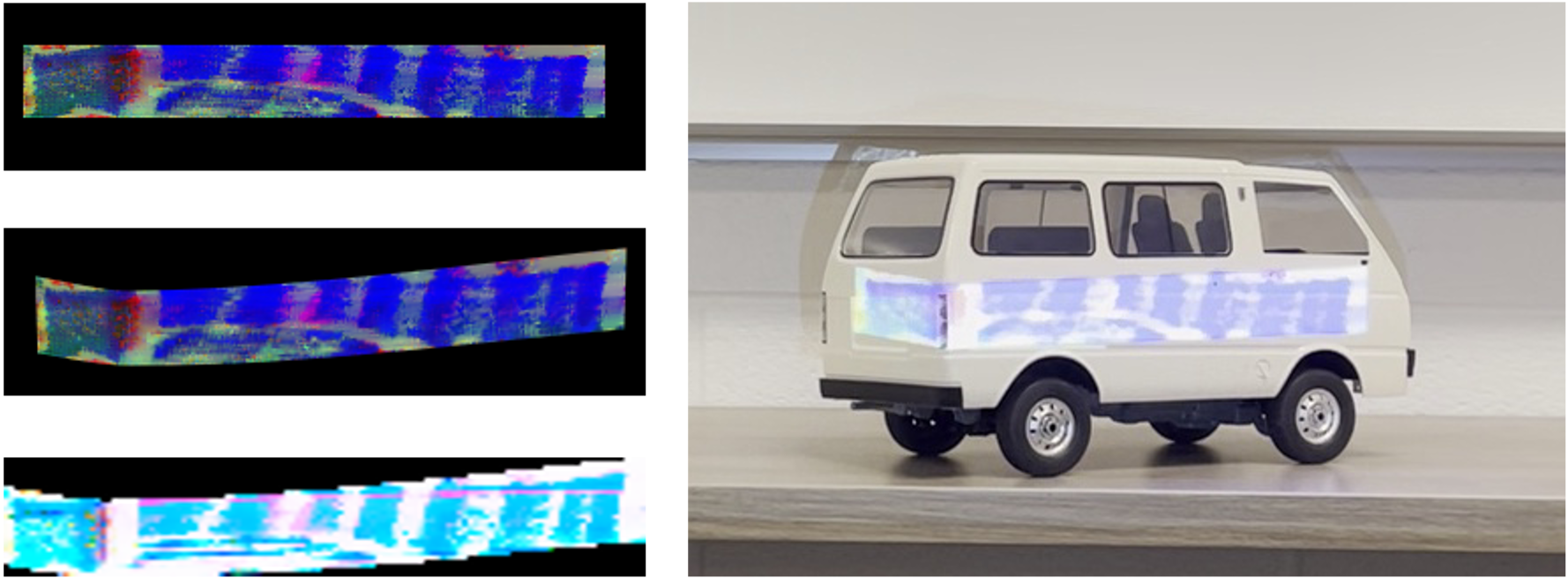}} 
    \subfigure[Attack on Mask R-CNN]{\includegraphics[width=0.48\textwidth]{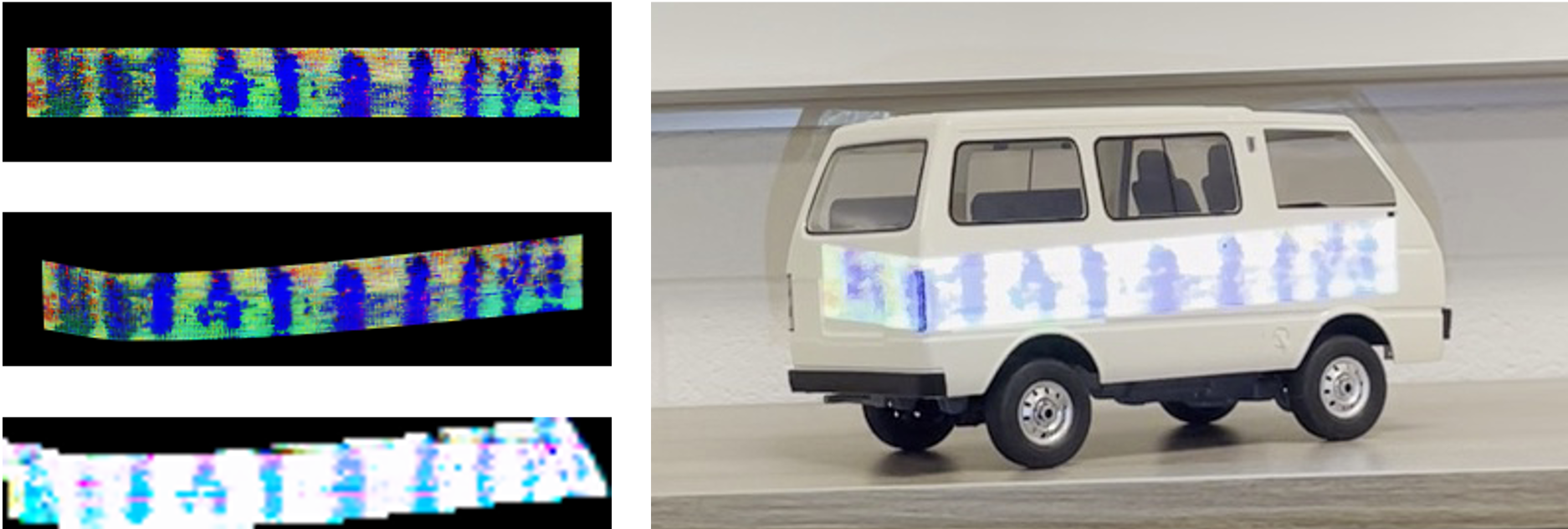}} 
    \caption{Visualization examples of attacks on Yolov3 and Mask R-CNN. The left column on both (a) and (b) is the well-trained projected attack patch, the simulated patch after geometric transformation, and the simulated patch on the vehicle from top to bottom. On the right side of each figure is the captured vehicle with the projected patch.}
    \label{fig:attack_examples}
\end{figure}

\subsection{Visualization of the Geometric Transformation}

Fig.~\ref{fig:TPS_visualization} illustrates an example of geometric transformation using TPS. Figs.~\ref{fig:TPS_visualization}(a)(d) are used to get source and target control points for TPS, respectively. We can then build up the geometric transformation model and obtain its corresponding parameters. Figs.~\ref{fig:TPS_visualization}(b)(c) are the simulated projected patch after geometric transformation, and the simulated projected patch on the vehicle. It can be seen that the simulated projected patch on the vehicle (Fig.~\ref{fig:TPS_visualization}(c)) and the data collected in real life (Fig.~\ref{fig:TPS_visualization}(d)) match very well. We utilize the parameters in the geometric transformation model and the color mapping model to convert the adversarial patch in the projector (Fig.~\ref{fig:TPS_visualization}(e)) to the adversarial patch on the vehicle (Fig.~\ref{fig:TPS_visualization}(f)). We add the transformed adversarial patch (Fig.~\ref{fig:TPS_visualization}(f)) onto the benign object image (Fig.~\ref{fig:TPS_visualization}(g)) to generate the final projected patch on the vehicle, which becomes the input to the object detector (Fig.~\ref{fig:TPS_visualization}(h)).

We also show visualization examples of attacks on Yolov3 and Mask R-CNN in Fig.~\ref{fig:attack_examples}. 
The left column on both Fig.~\ref{fig:attack_examples}(a) and Fig.~\ref{fig:attack_examples}(b) is the well-trained projected attack patch, the simulated patch after geometric transformation, and the simulated patch on the vehicle from top to bottom. The right columns are the captured vehicle with the projected patch in the physical world. It can be seen that the simulated patch on the vehicle and the physical world match well regarding the the patch shape and color.

\subsection{Attack Performance under Different Settings}

\begin{figure}[t]
    \centering    
    \subfigure[No attack under 100lux]{\includegraphics[width=0.32\textwidth]{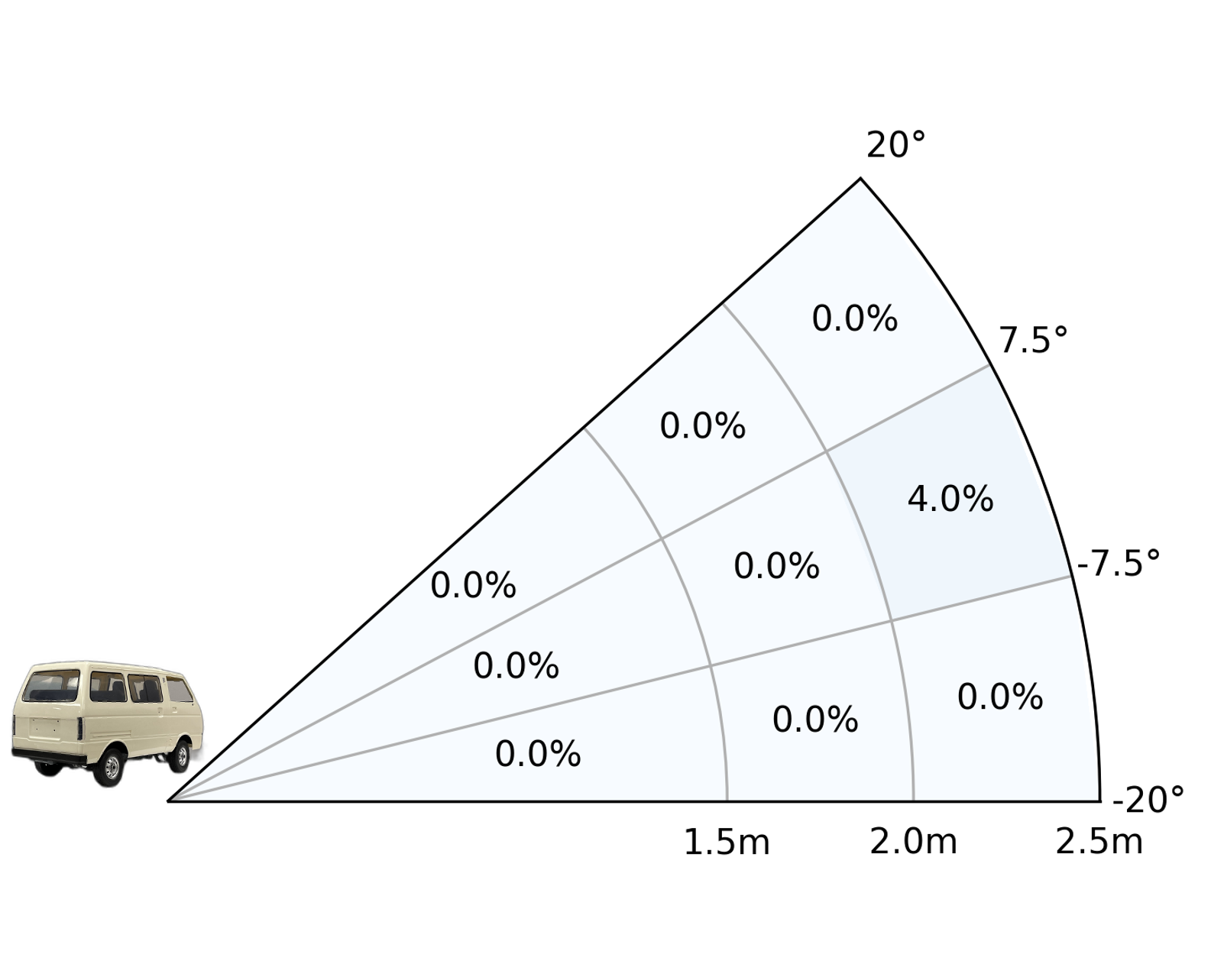}} 
    \subfigure[No attack under 200lux]{\includegraphics[width=0.32\textwidth]{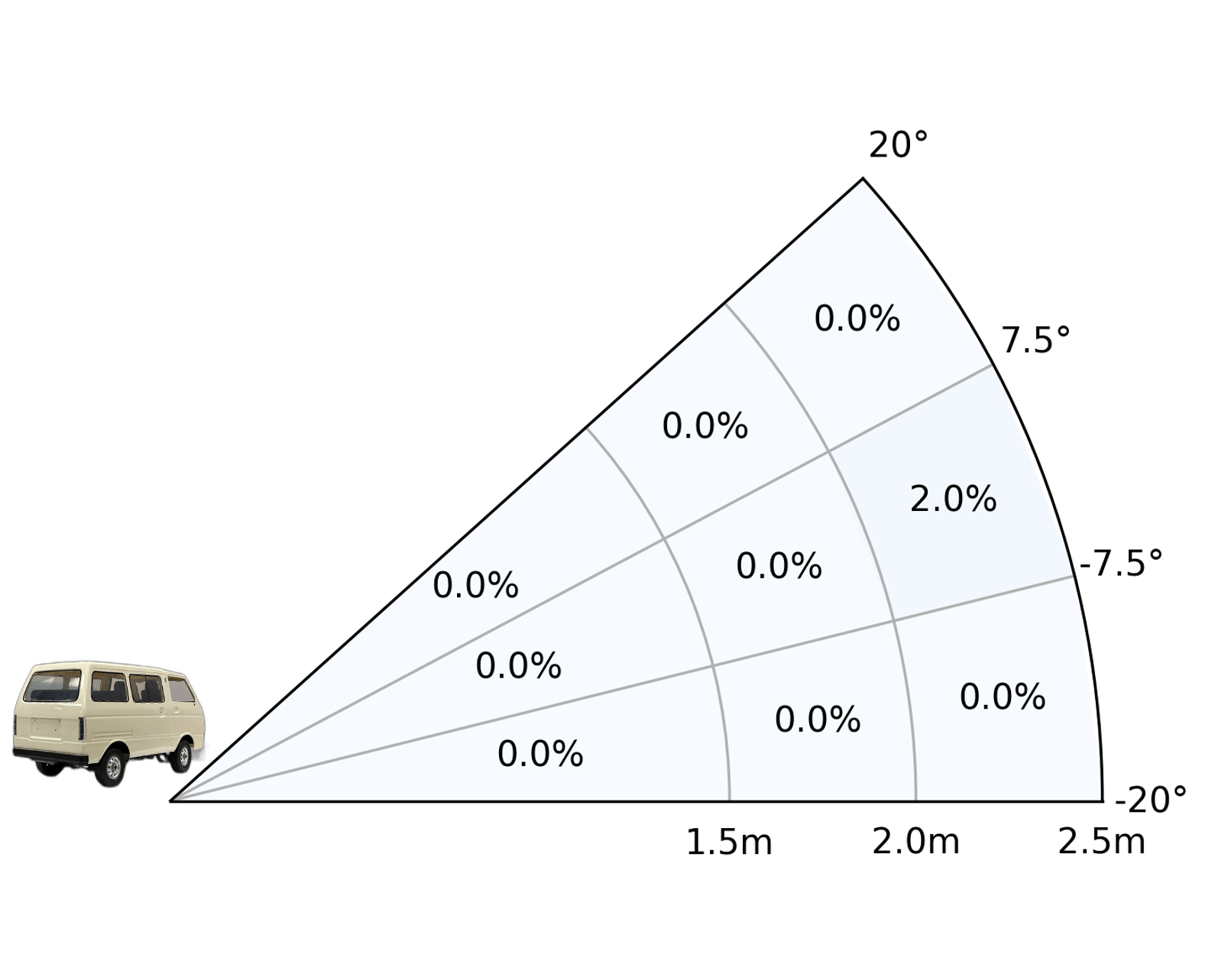}} 
    \subfigure[No attack under 500lux]{\includegraphics[width=0.32\textwidth]{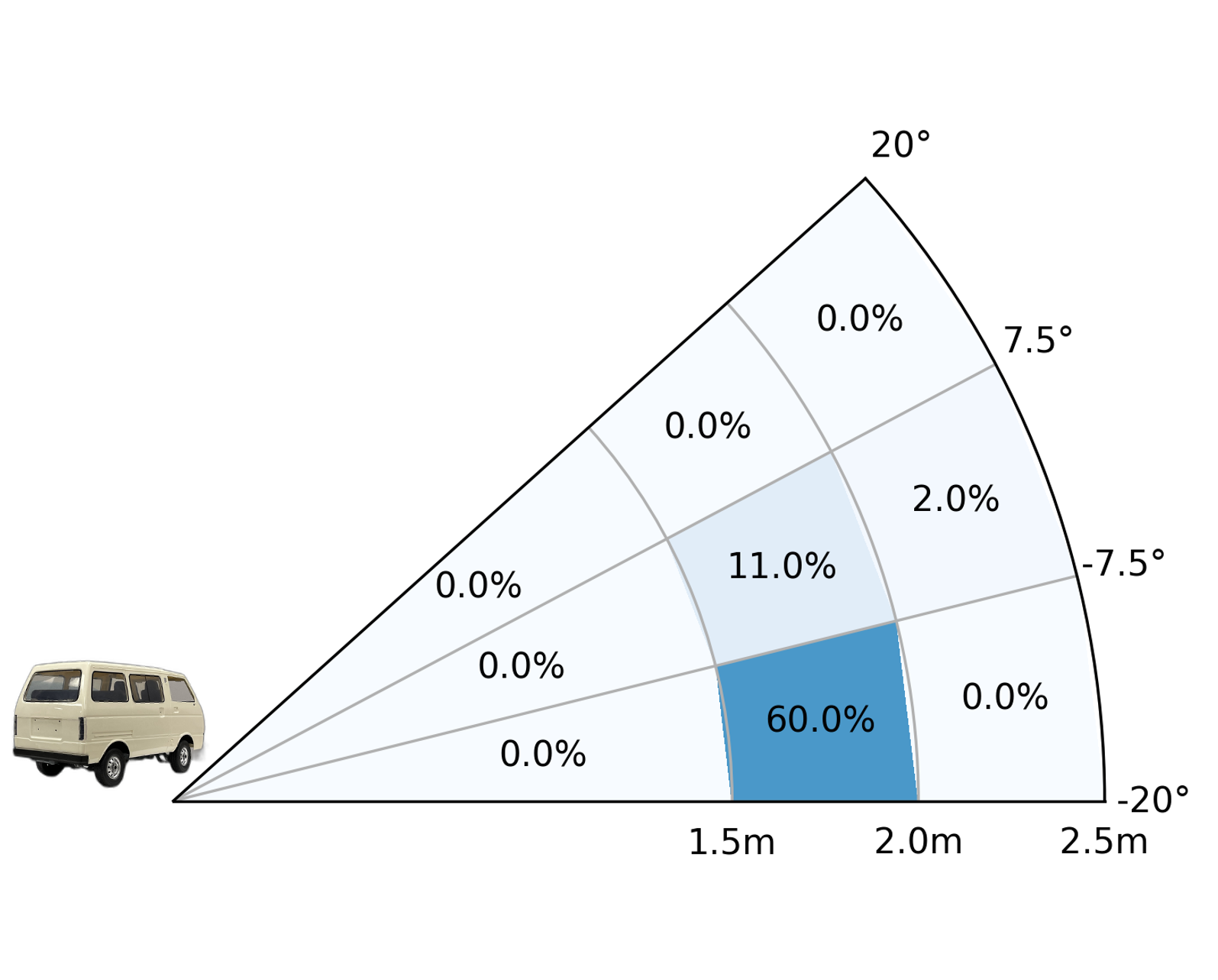}}   
    
    \subfigure[With attack under 100lux]{\includegraphics[width=0.32\textwidth]{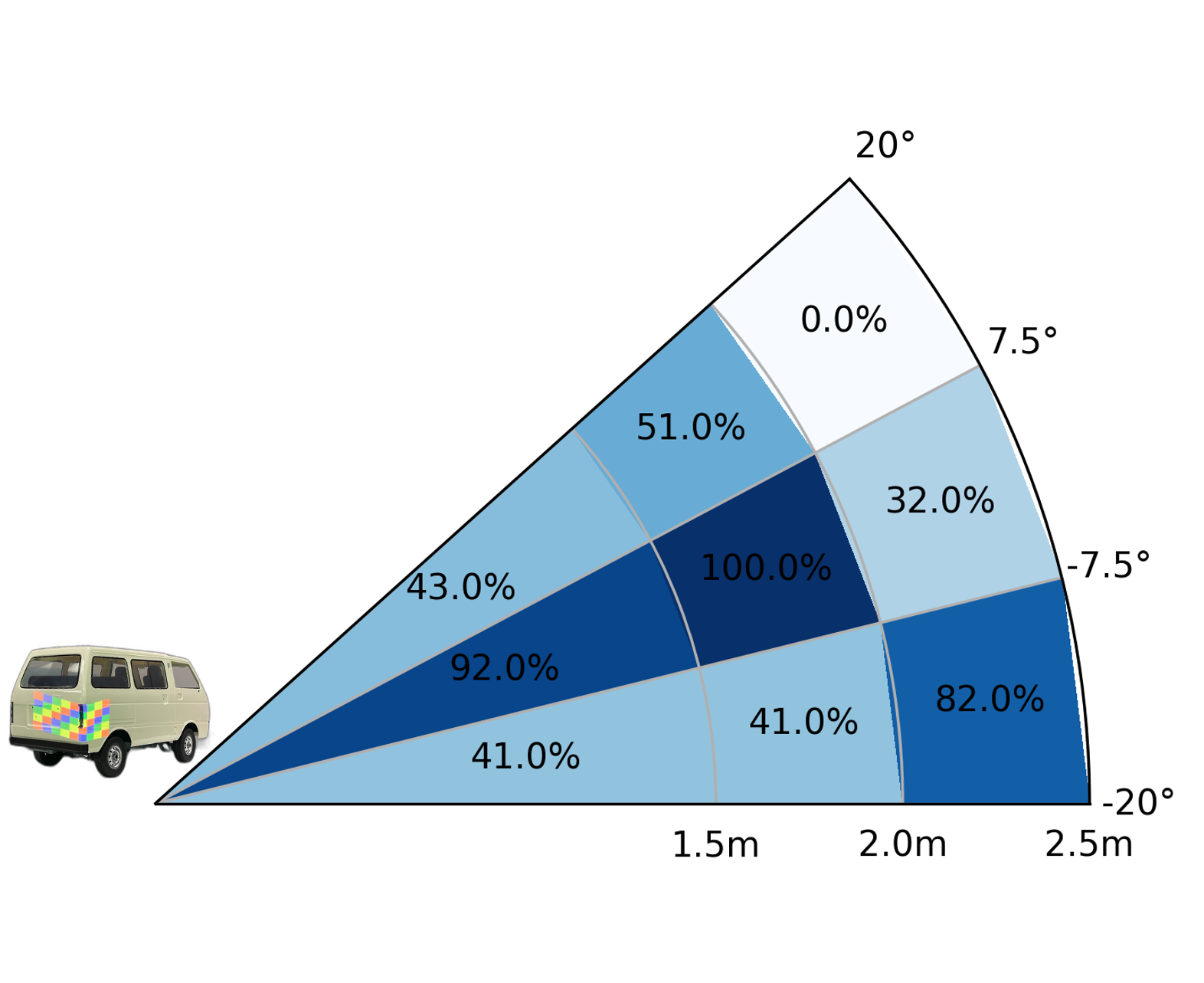}} 
    \subfigure[With attack under 200lux]{\includegraphics[width=0.32\textwidth]{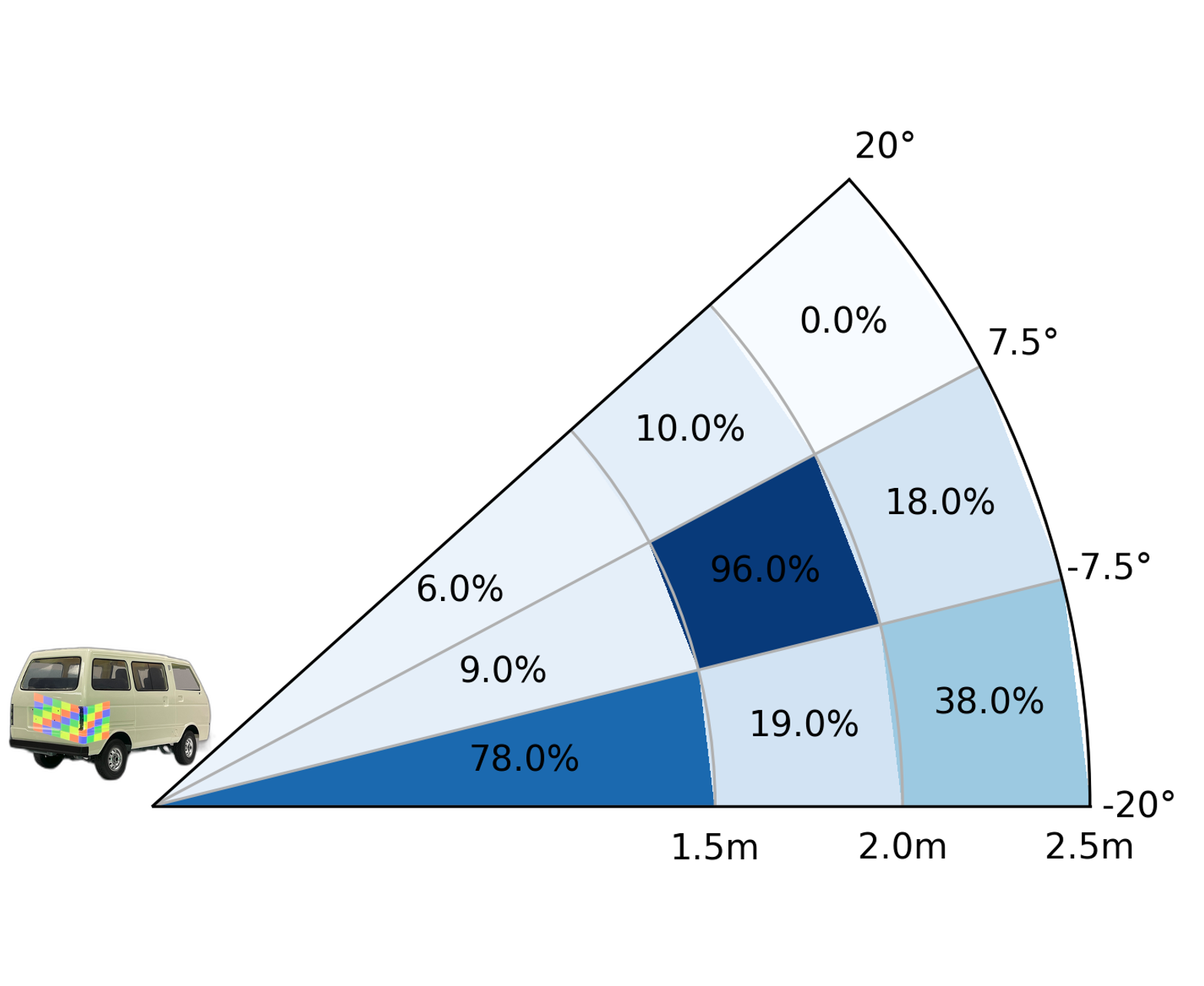}} 
    \subfigure[With attack under 500lux]{\includegraphics[width=0.32\textwidth]{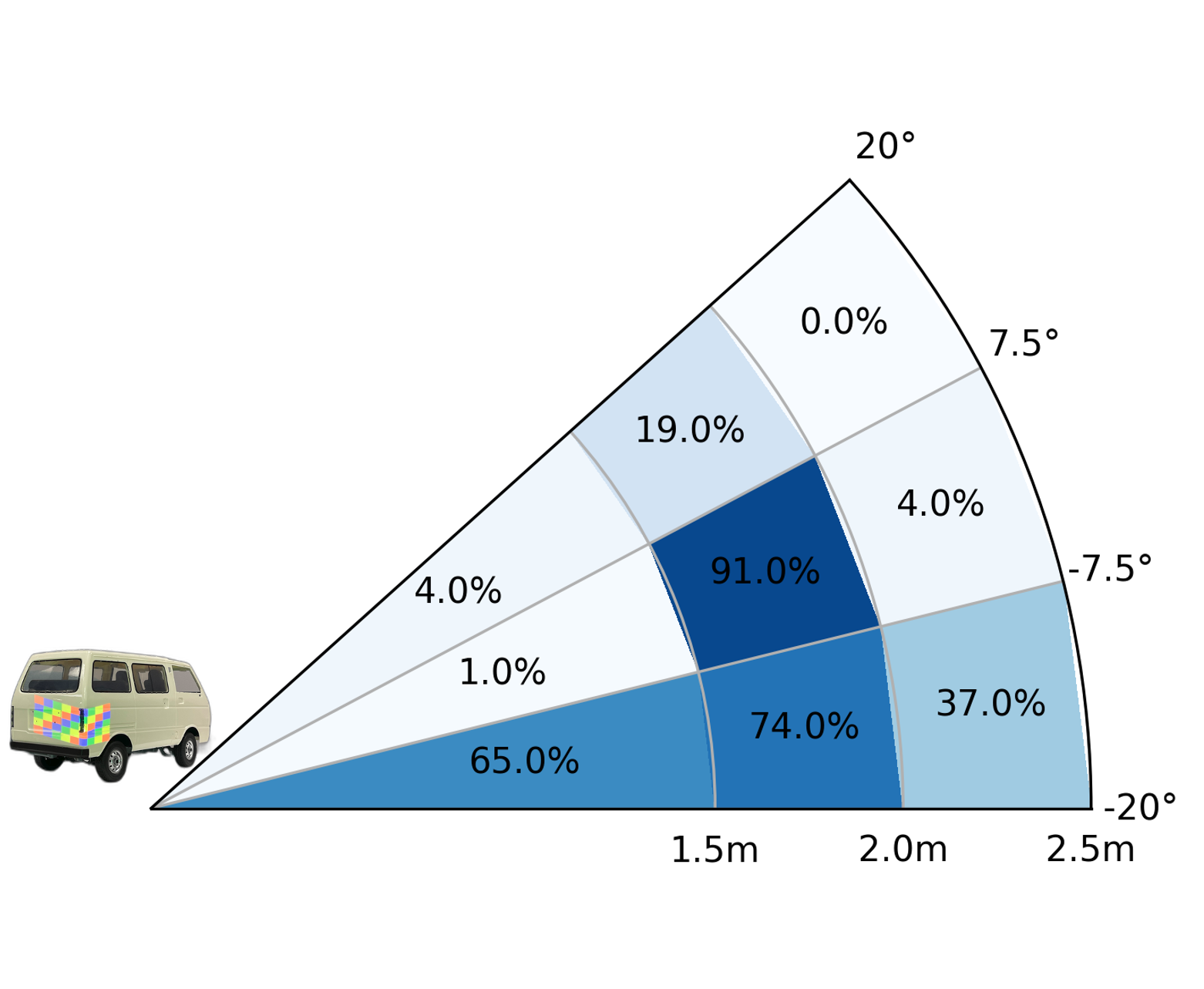}} 
    \caption{OMDR for Yolov3 with varying attack distance and attack angles under ambient light 100lux, 200lux and 500lux respectively. (a)(b)(c) show the OMDR without attack. (d)(e)(f) show the OMDR of 3D projection attack. }
    \label{fig:yolov3_attack_results}
\end{figure}

We evaluate the 3D projection attack under three ambient light conditions with varying distances and angles for Yolov3 as shown in Fig.~\ref{fig:yolov3_attack_results}. The first row shows the baseline results of Yolov3. When there is no attack present, OMDR is usually very close to 0. Only when the ambient light is 500 lux and the viewing angle is between $-20^\circ$ to $7.5^\circ$ at the attack distance of 2 meters, OMDR is relatively higher. The results might be attributed to the fact that the back view of the vehicle becomes harder to detect in Yolov3 under strong ambient light.  

For the scenarios with the attacks (Figs.~\ref{fig:yolov3_attack_results}(d)(e)(f)), it shows that OMDR significantly increases. With lower ambient light, the attack is usually more successful. This is because, as ambient light increases, the range of achievable colors diminishes due to the reduced impact of the projector-emitted light on the vehicle's appearance. Besides, when the attack distance is between 1.5 to 2 meters and the attack angle is between $-7.5^\circ$ to $7.5^\circ$, the average OMDR is around 96\%. Notably, when the ambient light is 100lux, we can achieve 100\% OMDR in this attack setting. A video demo of this attack can be viewed through the link: \url{https://youtu.be/8RbDpAAmsjs}.

Moreover, with the attack angle between $-20^\circ$ to $7.5^\circ$, it is usually easy to make the vehicle disappear in the detector. It is because that when the attack angle is between $7.5^\circ$ to $20^\circ$, part of the attack patch on the backside of the vehicle has less portion of the view as shown in Fig.~\ref{fig:different_angle}(c). On the other hand, the attack patch has a better view in the rest of the viewing angles in our attack setting (Fig.~\ref{fig:different_angle}(b)).

We also examine the performance of our attack on different object detectors under varying ambient light conditions at an attack distance of 1.5 meters. The average OMDR results for YOLOv3 and Mask R-CNN are summarized in Table~\ref{table:misdetection}. Our findings indicate that as ambient light increases, the success rate of the attack generally decreases for both models. Furthermore, Mask R-CNN consistently demonstrates greater resilience compared to YOLOv3, likely due to its ability to learn more robust features and its utilization of a region proposal network for detection~\cite{he2017mask}. However, this extra resilience comes at the expense of slower execution. For example, Mask R-CNN can take up to 14 times longer to perform than YOLOv3~\cite{lovisotto2021slap}.

%average adversarial computational time,  min max and average

% \usepackage{multirow}
% \usepackage{booktabs}
\begin{table}[t]
\centering
\begin{tabular}{cccccc}
\toprule
\multirow{2}{*}{\begin{tabular}[c]{@{}c@{}}Ambient Light \\ (lux)\end{tabular}} & \multicolumn{2}{c}{Yolov3}           & \multicolumn{2}{c}{Mask R-CNN}         \\ \cmidrule(r){2-3} \cmidrule(l){4-5} 
                                                                                & With Attack & W/O Attack & With Attack & W/O Attack \\ \midrule
100                                                                             & 59\%        & 0\%            & 46\%        & 26\%           \\
200                                                                             & 31\%        & 0\%            & 36\%        & 30\%           \\
500                                                                             & 23\%        & 0\%            & 18\%        & 25\%           \\ \bottomrule
\end{tabular}
\caption{Average misdetection rate under different ambient light conditions}
\label{table:misdetection}
\end{table}

%\subsection{Attack Tansferability}
%show attack transferability for SSD yolov5 yolov8 using the data of 100lux ambient light 

\section{Discussion}
We mainly discuss the feasibility and practicality of the 3D projection attack.

\subsubsection{Attack Feasibility and Practicality.}
Unlike patch attacks, which involve placing physical stickers or patches on objects, projection attacks can be temporary. The projections can be turned on and off quickly, making them harder to detect and track over time. Once the projection is turned off, there are no physical traces left behind, unlike patches or stickers that can be discovered upon inspection.

Projection attacks can change patterns dynamically, adapting to different conditions and object surfaces. This flexibility makes it more challenging for detection systems to recognize and filter out adversarial patterns. Projected adversarial patches can also be modified in real time based on the feedback from the detection system, which allows for continuous optimization of the attack and maintains its effectiveness.

\subsubsection{Ambient Light.} Our experiments show that the increasing ambient light could rapidly degrade the effectiveness of the attack in bright conditions. In a real-world physical attack scenario, the attacker could launch the attack during the daytime on overcast days or near sunset or sunrise when the ambient light level is below 400 lux.

\subsubsection{Moving Target Object.} Our attack is effective only on static objects or those moving at low speeds relative to the attacking device. Otherwise, targeting the vehicle becomes problematic. Therefore, our attack is more suitable for vehicles parked on the roadside or traffic cones on the road. Though the attack may only last for several seconds due to movement issues, it is a split-second attack. This means that even if it functions for only a few seconds, it can potentially lead to severe consequences.

\subsubsection{Data Collection.} The geometric transformation model and the color mapping model need to be well-trained in advance. Consequently, essential data collection is required. The attacker may purchase or locate an identical object to the target and gather all necessary data to develop these models beforehand to launch a more accurate attack.

\subsubsection{Potential Defense.}
One potential defense method is adversarial training~\cite{liu2022segment,goodfellow2014explaining}, which incorporates AEs, including those generated by projection attacks, into the training process. This helps the model learn to recognize and ignore projected adversarial patterns. Previous research~\cite{liu2022segment} proposed a method to locate and segment the adversarial patch, incorporate adversarial training, and then completely remove the patch. However, in the case of our projection attack, if the projected patch size on the target object is enlarged or if the patch fully covers the object, removing the patch entirely might still result in the benign object being mis-detected due to the large area of the object being obscured.

Another possible defense is to verify the consistency of detections over time. Since our projection attack is short-lived, it may cause sudden changes in detected objects over time. Checking for temporal consistency can help identify and discard adversarial detections and alert the human driver.

\section{Conclusion}

We propose a transient adversarial 3D projection attack targeting object detection in autonomous driving scenarios, which projects a well-crafted adversarial patch onto a 3D surface. The 3D projection attack is formulated as an optimization problem, combining a color mapping model and a geometric transformation model. We enhance the robustness of our attack by considering various environmental factors. We conduct experiments to evaluate the proposed attack against YOLOv3 and Mask R-CNN object detectors in physical attack scenarios. Our evaluation results show an attack success rate of up to 100\% under low ambient light conditions. This research underscores the need for defense strategies to mitigate transient projection attacks on AI-driven autonomous vehicles. Future work should focus on creating adaptive and resilient countermeasures that can detect and neutralize such attacks in real time, thereby safeguarding both passengers and pedestrians in diverse driving conditions.

%, highlighting the potentially severe consequences of the adversarial 3D projection attack in real-world driving scenarios.
\section*{Acknowledgements}

We would like to extend our appreciation to the anonymous reviewers for their invaluable input on our
study. This work was supported in part by the U.S. National
Science Foundation grant CNS-2235231.

\bibliographystyle{splncs04}
\bibliography{reference}

\end{document}